\documentclass[review]{elsarticle}

\usepackage{lineno,amssymb,graphicx}
\usepackage{amsmath}
\usepackage{bm}
\usepackage{epstopdf}%
\usepackage{bm}
\usepackage{multirow}
\usepackage{booktabs}
\modulolinenumbers[5]

\journal{Nuclear Instruments and Methods in Physics Research A}

 \newcommand{\muo}{ \mu_{\scriptscriptstyle 0}}

\begin{document}

\begin{frontmatter}

\title{Analytic models of magnetically enclosed spherical and solenoidal coils}

\author{C.-Y. Liu$^{a,b}$, T. Andalib$^c$, D.C.M. Ostapchuk$^a$, C.P. Bidinosti$^{a,c,*}$}
\address{$^a$Department of Physics, University of Winnipeg, Winnipeg, MB, R3B 2E9, Canada}
\address {$^b$Department of Electrical and Computer Engineering, University of Manitoba, Winnipeg, MB, R3T 5V6, Canada}
\address {$^c$Department of Physics and Astronomy, University of Manitoba, Winnipeg, MB, R3T 2N2, Canada}
\address {$^*$Corresponding authour, E-Mail address: c.bidinosti@uwinnipeg.ca}

\begin{abstract}
We provide analytic solutions of the net magnetic field generated by spherical  and solenoidal coils enclosed in highly-permeable, coaxial magnetic shields. We consider both spherical and cylindrical shields in the case of the spherical coil and only cylindrical shields for the solenoidal coil. 
Comparisons  of field homogeneity are made and we find that the solenoidal coil produces the more homogeneous field for a given number of windings. The models  are useful as theoretical and conceptual guides for coil design, as well as for benchmarking finite-element analysis. We also demonstrate how the models can be generalized to explore field inhomogeneities related to winding misplacement.
\end{abstract}

\begin{keyword}
Shield-coupled coils \sep  Spherical Coil \sep Solenoidal coil \sep Magnetic field homogeneity \sep Neutron electric dipole moment
\end{keyword}

\end{frontmatter}


\section{Introduction}

Our motivation for this work is  the development of a proposed experiment to measure the electric dipole moment (EDM) of the neutron 
at the new ultracold neutron facility at TRIUMF~\cite{TUCANucn,TUCANbeam}.  The observation of a permanent EDM of a particle or system  would provide a direct signal of time-reversal symmetry breaking, which in turn could help our understanding of important questions such as physics beyond the standard model or the cosmological matter-antimatter asymmetry~\cite{Chupp}.  EDM experiments are very challenging, however, especially in regard to their need  for  magnetic field homogeneity and stability. 
As a result, the typical approach is to (i) enclose the experiment inside  magnetic shielding in order to eliminate external field sources, and  (ii) construct internal current structures that generate the desired homogeneous measurement field~\cite{Chupp}.

\begin{figure}[htb]
\centering
\includegraphics[]{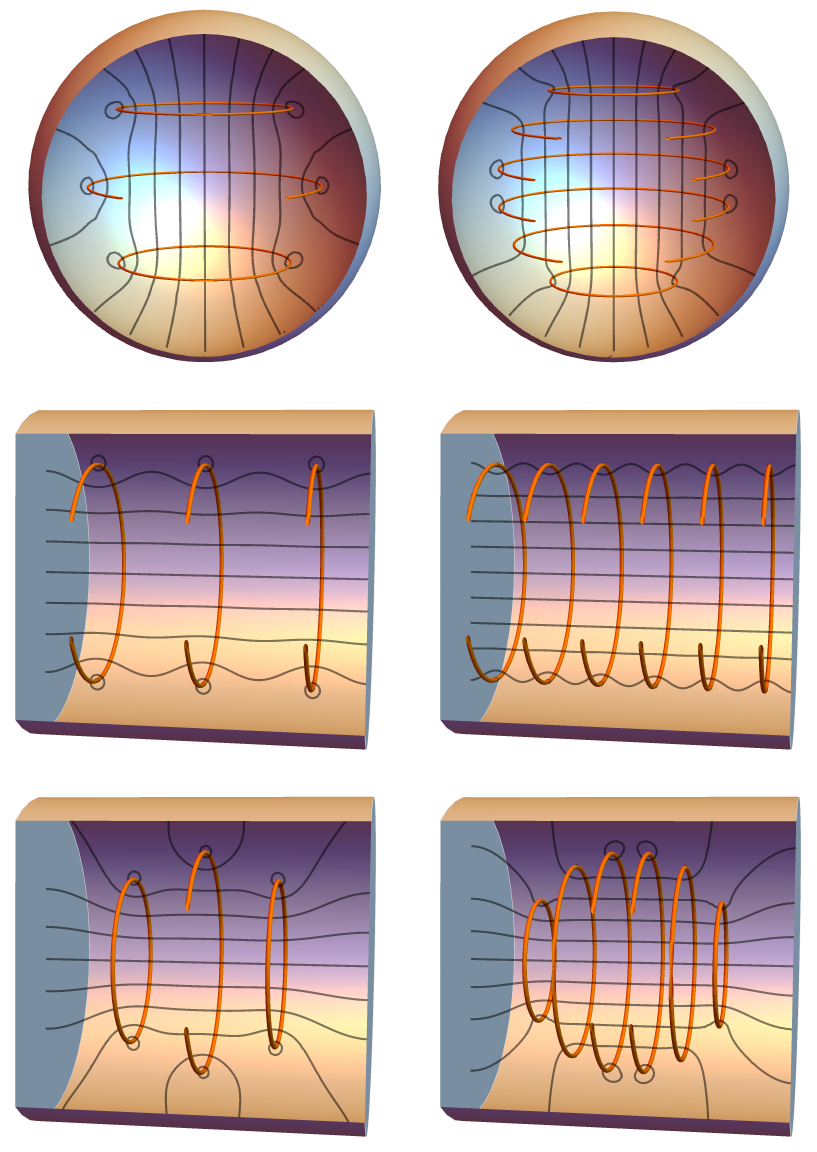}
\caption{Cutaway drawings of the spherical coil in a spherical shield (top) and the solenoidal coil (middle) and spherical coil (bottom) in cylindrical shields, with three (left) and six (right) current loops. The field lines -- which are contours of $\rho A_\phi$~\cite{haus} calculated from the formulae for the vector potential provided here -- enter the high-$\mu$ shields at normal incidence as expected.}
\label{Blines}
\end{figure}

There are a number of well-known surface current distributions $\bm F(\bm r)$ that generate perfectly uniform magnetic fields, most notably the following: (i) $F \sin\theta \, \bm{\hat{\phi}}$ on a  spherical surface, (ii) $F \, \bm{\hat{\phi}}$ on an infinitely-long cylindrical surface, and (iii) $F \sin\phi \, \bm{\hat{z}}$ on an infinitely-long cylindrical surface.  A comparison of the magnetic field uniformities of  finite-sized, discrete-current  approximations of these three surface currents was recently presented~\cite{nouri}.  In this work, we consider the effect of a highly-permeable magnetic shield on the field generated by an internal coil structure, a scenario more germane to the typical EDM experiment.  In particular, we provide general analytic solutions for the axi-symmetric cases of the spherical and solenoidal coils inside fully enclosed co-axial shields as shown in Fig.~\ref{Blines}.  For simplicity, we consider the shields to have very high linear permeability, \textit{i.e.}, $\mu\gg\muo$. 

The paper is organized as follows. In Sections~\ref{Sec:SphericalCurrents} and~\ref{sec:CylindricalCurrents}, formula for the magnetic field of the spherical coil inside a spherical shield and the solenoidal coil inside a cylindrical shield are derived and explored.  In Section~\ref{sec:SphericalCoilCylindricalShield}, we discuss how the preceding results can be used to analyze the spherical coil inside a cylindrical shield, a configuration that has been proposed for a neutron EDM experiment~\cite{masuda}.  In Section~\ref{sec:Comparison}, a comparison of  magnetic field maps of the three coil types is presented and we find that the solenoidal coil produces the more homogeneous field for a given number of windings. In Section~\ref{sec:WindingMisplacement}, we explore the impact of winding misplacement on the homogeneity of the solenoidal coil.   Finally, appendices are provided to give further details of the derivations and results discussed in the main text.

\clearpage

  \section{The Spherical coil inside a spherical shield}  
  \label{Sec:SphericalCurrents}

We begin by considering the general zonal surface current $\bm F = F_\phi(\theta)   \, \bm{\hat{\phi}}$  on a sphere of radius $a$.  The resulting vector potential is~\cite{smythe} 
\begin{equation}
 A_\phi(r,\theta) = -\muo \, \sum_{n=1}^\infty \frac{C_n P_n^1(u)}{2n+1}  
\begin{cases}
(r/a)^n  &  r < a    \\[.1cm]
(a/r)^{n+1} &  r > a 
\end{cases}
\, ,
\label{Asphere}
\end{equation}
where $P_n^1(u) $ is the associated Legendre function\footnote{In this and previous work~\cite{bidAIP}, we observe Smythe's definition of $P_n^m(u)$  -- see Eq.(6), \S5.23 in Ref.~\cite{smythe} -- which does not include the Condon-Shortley phase $(-1)^m$.}
 of order 1 and
degree $n$, $u=\cos\theta$, and the coefficients are
\begin{equation}
C_n 
=\frac{-a(2n+1)}{2n(n+1)} \, \int_{0}^{\pi} F_\phi(\theta)\,  P_n^1(\cos \theta) \, \sin \theta \, {\rm d}\theta \, .
\label{Cn}
\end{equation}
General expressions for the  magnetic field components $ B_r $ and $B_\theta$ 
resulting from  Eq.~\ref{Asphere} are found in Ref.~\cite{smythe} and  not reproduced here.

To include the response of a spherical shield of linear permeable material with inner radius $b\ge a$ and outer radius $c$, one may solve the equivalent problem of bound surface currents on $r=b$ and~$c$ subject to the appropriate boundary conditions~\cite{bidAIP, bidIEEE}.  
For this work, however, we are interested only in high-$\mu$ shields (i.e. $\mu\gg\muo$), and as a result  one can take advantage of the fact that the magnetic field must enter the shield at normal incidence at $r=b^-$ as $\mu \rightarrow \infty$, regardless the value of $c$.  It suffices, then, to consider here a single bound current  on $r=b$ only (effectively letting $c \rightarrow \infty$.)  
In doing so, the problem simplifies greatly and it is straightforward  to satisfy the boundary condition $B_\theta(b^-)=0$ for the net field generated by the surface currents on $r=a$ and~$b$.  In the end, one finds that the vector potential due to $F_\phi(\theta)$ inside a high-$\mu$ spherical shield is independent of the shield thickness and given by
\begin{equation}
 A_\phi(r,\theta) = -\muo \, \sum_{n=1}^\infty \frac{C_n P_n^1(u)}{2n+1}  
\begin{cases}
(\tfrac{r}{a})^n \left[ 1 + \tfrac{n}{n+1}\, (\tfrac{a}{b})^{2n+1} \right] &  r < a    \\[.1cm]
(\tfrac{a}{r})^{n+1} \left[ 1 + \tfrac{n}{n+1}\, (\tfrac{r}{b})^{2n+1} \right]  &  a<r <b 
\end{cases}
\, .
\label{AsphereShield}
\end{equation}
In the region of interest, $r<a$, the components of the magnetic field are
\begin{eqnarray}
\left( \!\!\!
\begin{array}{c}
B_r \\ 
B_\theta
\end{array}
\!\!\! \right) 
&=& \frac{\muo }{a} \, \sum_{n=1}^\infty \left[ 1 + \frac{n}{n+1}\, \left(\frac{a}{b}\right)^{2n+1} \right]  \nonumber \\
&&  \times \frac{ C_n (n+1)}{2n+1}  \, \left(\frac{r}{a} \right)^{n-1}
\left( \!\!\!
\begin{array}{c}
-n P_n(u)  \\ 
P_n^1(u)
\end{array}
\!\!\! \right) \, ,
\label{BsphereShield}
\end{eqnarray}
where $P_n(u)$ is the Legendre function of degree $n$ and the term in square braces, known as the reaction factor~\cite{bidAIP,urankar},  quantifies the extent to which the $n$-th term in the field expansion is augmented by the presence of the shield.  An important consequence of the reaction factor, discussed further below,  is that the homogeneity of the field of a spherical coil may  either improve or degrade inside a shield depending on the ratio $a/b$~\cite{bidAIP}.

\subsection{The sine-theta surface current}

A surface current of  the form $F_\phi(\theta) = F \, \sin\theta$, comprising the $n=1$ harmonic only, generates a perfectly homogeneous magnetic field in the region $r<a$, whether it  lies in free space or is surrounded by a spherical shield of any thickness and  linear permeability~\cite{bidAIP}.    For the case of a high-$\mu$ shield, one finds from Eqs.~\ref{Cn} and~\ref{BsphereShield} that the internal magnetic field is
\begin{equation}
\left( \!\!\!
\begin{array}{c}
B_r \\ 
B_\theta
\end{array}
\!\!\! \right) 
= \frac{2\muo F}{3}  \left[ 1 +  \frac{a^3}{2  b^3}  
\right]
\left( \!\!\!
\begin{array}{c}
\cos \theta  \\ 
-\sin \theta
\end{array}
\!\!\! \right) \, .
\label{Bsinetheta1}
\end{equation}
The corresponding cylindrical  components of the field are  
\begin{equation}
\left( \!\!\!
\begin{array}{c}
B_\rho \\ 
B_z
\end{array}
\!\!\! \right) 
= \frac{2\muo F}{3} \left[ 1 +  \frac{a^3}{2  b^3} 
 \right]
\left( \!\!\!
\begin{array}{c}
0\\ 
1
\end{array}
\!\!\! \right) \, ,
\label{Bsinetheta2}
\end{equation}
where $B_\rho= B_r  \sin\theta+ B_\theta \cos\theta $
and  $B_z= B_r  \cos\theta- B_\theta \sin\theta $.  These provide a more natural description of the  axisymmetric fields encountered here.

\subsection{The spherical coil}
\label{sphericalcoil}
\begin{figure}[htb]
\centering
\includegraphics[trim= 0 0 0 0, clip=true,keepaspectratio, width=3.2in]{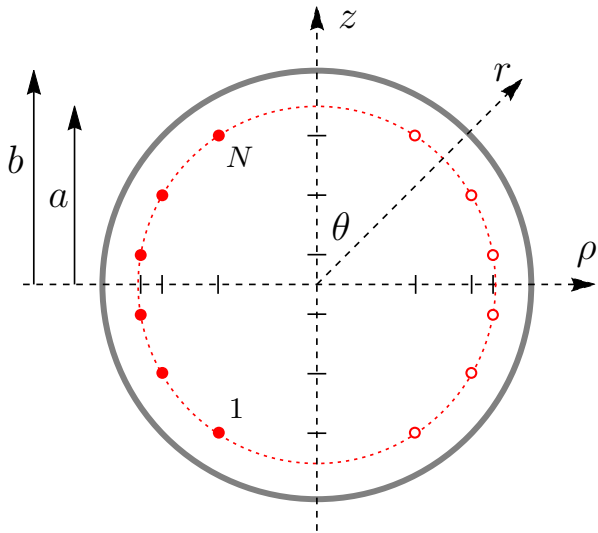}
\caption{A cross-sectional view of the  $N$-loop spherical coil  on radius $a$ (dashed red line) inside a high-$\mu$ shield of inner radius $b$ (solid gray line). Standard spherical and cylindrical coordinates are employed with  $\rho=r\sin\theta$ and $z=r \cos\theta$.  The hash marks on the axes indicate the axial and radial coordinates $(\rho_i,z_i)$ of the $N=6$ current loops depicted here.  The solid versus open circles indicate current flow out of and into the page, respectively.}
\label{geometrySphere}
\end{figure}

The sine-theta surface current of the preceding section comprises a total current
\begin{equation}
I_{\rm tot}=  
\int_0^\pi F \sin \theta \, a \,{\rm d}\theta= \int_{-a}^a F  \,{\rm d}z = 2aF \, ,
\label{Itot}
\end{equation}
and can be approximated by $N$ discrete loops, each carrying current $\delta I =2aF/N$ and mutually separated  by axial distance $\delta z = 2a/N$~\cite{haus}. 
 A suitable scheme for constructing such a 
 spherical coil is to place a current loop at the mid-point of each increment $\delta z$~\cite{nouri}.  In ascending order along the $z$-axis, then, the $i$-th loop is located at $z_i =-a+\delta z \,(i-\tfrac{1}{2})=a(-1 +\tfrac{2i-1}{N})$, as shown in Fig.~\ref{geometrySphere}.

A discrete loop carrying current $I$ and located at polar angle $\theta_i$ on $r=a$ can be treated as the surface current $F_\phi(\theta) = \tfrac{I}{a} \, \delta(\theta-\theta_i)$ in spherical coordinates.  The field generated by such a loop in the region $r<a$ is given by Eq.~\ref{BsphereShield} with coefficients~\cite{smythe}
\begin{equation}
C_n= - \frac{I(2n+1)}{2n(n+1)} \, \sin\theta_i \, P_n^1(\cos\theta_i)
\label{CnLoop}
\end{equation}
from Eq.~\ref{Cn}.
The net  field of the spherical coil, then, is the sum of contributions  from each loop $i$, with $\cos \theta_i=z_i/a$ and $\sin \theta_i=\rho_i/a=\sqrt{1-(z_i/a)^2}$.  
For $r<a$, the result is 
\begin{equation}
\left( \!\!\!
\begin{array}{c}
B_r \\ 
B_\theta
\end{array}
\!\!\! \right) 
=  \frac{\muo I}{ 2 a}  
\sum_{n=1,3,5,\dots}^\infty   \! b_n \left(\frac{r}{a}\right)^{\!n-1}  \,
\left( \!\!\!
\begin{array}{c}
 P_n(\cos \theta)  \\ 
-\tfrac{1}{n}P_n^1(\cos \theta)
\end{array}
\!\!\! \right) \, ,
\label{BNloop}
\end{equation}
where  only odd values of $n$ survive by symmetry, and
\begin{equation}
b_n = \left[ 1+\frac{n}{n+1} \,  
\left(\frac{1}{\beta} \right)^{2n+1} 
\right] \, \sum_{i=1}^N \,  \varrho_i \,  P_n^1(\zeta_i) \, ,
\label{bn}
\end{equation}
with dimensionless parameters $\varrho_i=\rho_i/a$, $\zeta_i=z_i/a$,
and $\beta=b/a$.  Both the number of loops $N$ and the ratio of the shield to coil radii $\beta$ have a strong  influence on the uniformity of the magnetic field of the spherical coil, as will be shown below. As $N\rightarrow \infty$, the summation in Eq.~\ref{bn} goes to zero for all values of $n$ except  $n=1$, in which case it equals $2N/3$ and one recovers 
the perfectly uniform internal field of Eq.~\ref{Bsinetheta1} for  a  sine-theta surface current with $I_{\rm tot}=NI$ and $F=NI/2a$.

Making use of recurrence relations~\cite{smythe}, the cylindrical components of the magnetic field of the spherical coil can be written as
\begin{eqnarray}
B_\rho &=& \frac{-\muo I}{ 2 a}  \sum_{n=3,5,\dots}^\infty   \! b_n \left(\frac{r}{a}\right)^{\!n-1}  \frac{\sin\theta}{n} \, P'_{n-1}(\cos \theta)  \nonumber \\
&=&  \frac{\muo I}{ 2 a} \left(- b_3 (\varrho \zeta) + b_5 (\tfrac{3}{2}\varrho^3\zeta -2\varrho\zeta^3)+  
\dots \right)  \nonumber 
\\
&=& \frac{\muo I}{ 2 a}  \sum_{n=3,5,\dots}^\infty   \!\! b_n  \, \mathcal{P}_n(\varrho,\zeta)  \label{BrhoSpherical}  \\[.4cm]
B_z &=& \frac{\muo I}{ 2 a}  \sum_{n=1,3,5,\dots}^\infty   \! b_n  \left(\frac{r}{a}\right)^{\!n-1} P_{n-1}(\cos \theta) \nonumber  \\
&=&  \frac{\muo I}{ 2 a} \left(b_1- b_3(\tfrac{1}{2}\varrho^2-\zeta^2) +\dots \right)  \nonumber \\
&=& \frac{\muo I}{ 2 a} \left(b_1+ \sum_{n=3,5,\dots}^\infty   \!\! b_n   \, \mathcal{Q}_n(\varrho,\zeta) \right) \label{BzSpherical}    \, ,
\end{eqnarray}
where 
 $\varrho=\rho/a$ and $\zeta =z/a$ are dimensionless coordinates, and the polynomials $\mathcal{P}_n$ and  $\mathcal{Q}_n$ 
 (described further in~\ref{appendixA}) comprise terms of the form $\varrho^{n-1-l}\zeta^l$ with $l=0,1,\dots, n-1$.
Because the  magnetic field  is axisymmetric, the same coefficient $b_n$ appears in all terms of spatial order $n-1$. 
Dividing Eqs.~\ref{BrhoSpherical}  and~\ref{BzSpherical} by the value of the uniform field term $B_z(0,0)=\muo I b_1/2a$ gives
\begin{eqnarray}
\frac{B_\rho(\rho,\zeta)}{B_z(0,0)}& =&  \sum_{n=3,5,\dots}^\infty   \!\! \frac{b_n}{b_1}  \, \mathcal{P}_n(\varrho,\zeta) \label{BrhoSphericalNorm} \\
\frac{B_z(\rho,\zeta)}{B_z(0,0)}& =& 1 +  \sum_{n=3,5,\dots}^\infty   \!\! \frac{b_n}{b_1}  \, \mathcal{Q}_n(\varrho,\zeta) \, ,
\label{BzSphericalNorm} 
\end{eqnarray}
and one can use the normalized coefficients $b_n/b_1$ to characterize and compare the 
 inhomogeneities of different coil configurations over the entire  inner region $\sqrt{\varrho^2+\zeta^2}<1$.

\begin{figure}[htb]
\centering
\includegraphics[trim= 0 0 0 0, clip=true,keepaspectratio, width=7.6cm]{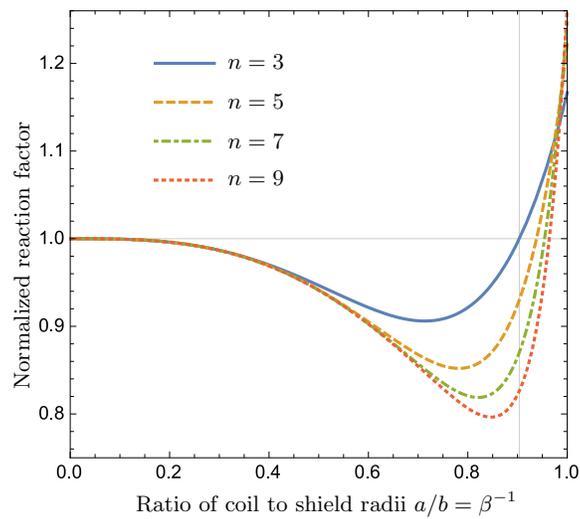}
\caption{The normalized reaction factor versus $\beta^{-1}$ 
shown for $n=3, 5, 7$ and 9. For any given number of loops $N$, the normalized reaction factor for every degree $n$ is less than 1.0 over the range $0<\beta^{-1}<0.9036$ (vertical gray line). This tends to improve the homogeneity of any spherical coil  with $a< 0.9036\, b$.
However, as $\beta^{-1} \rightarrow 1$, all normalized reaction factors are greater than 1.0, which tends to degrade the homogeneity of any spherical coil  of finite turns inside a tight-fitting shield ($b=a$). }
\label{SphericalReactionFactors}
\end{figure}

\begin{figure}[htb]
\centering
\includegraphics[keepaspectratio, width=8.3cm]{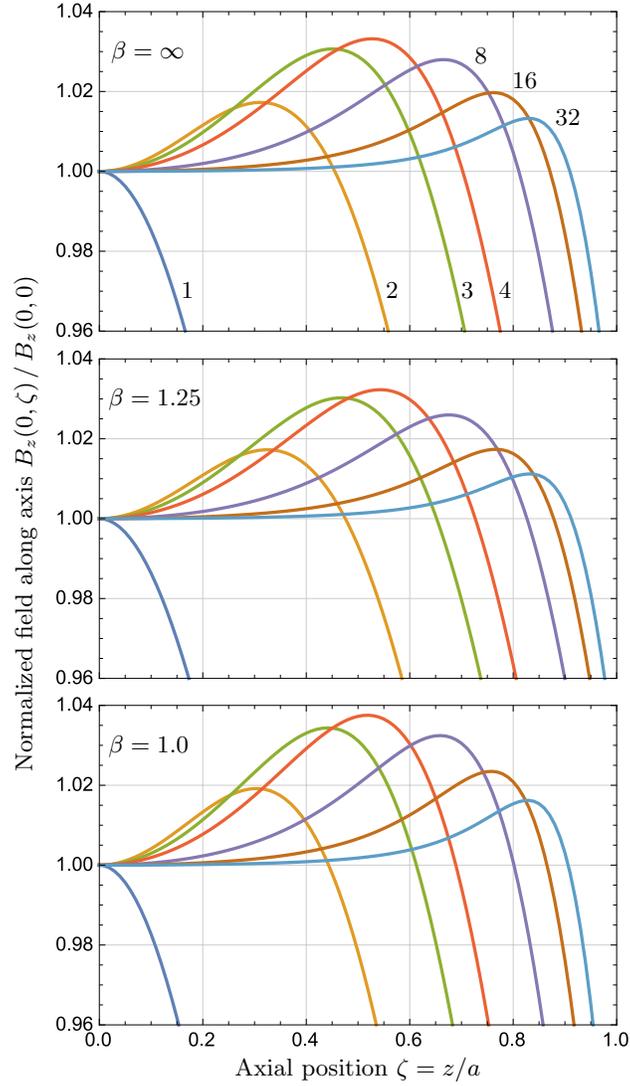}
\caption{The normalized axial field of the spherical coil in free space (top), in a shield of inner radius $b=1.25a$ (middle), and in a tight-fitting shield (bottom).  The homogeneity improves with the number of loops $N$ (plots are labeled in the top graph and color-coded throughout).   For a given $N$, peak deviations from the central field are smallest for the middle configuration ($\beta^{-1}=0.8$), while they are largest for the bottom configuration. }
\label{Bzplot}
\end{figure}

From Eq.~\ref{bn}, 
the normalized coefficients are  
\begin{equation}
\frac{b_n}{b_1}= 
\left[ 
\frac{\beta^3 +\tfrac{n}{n+1}\beta^{-2n+2} }{\beta^3+\tfrac{1}{2}}
\right] 
 \times  
\left(
\frac{\sum_{i=1}^N \,  \varrho_i \,  P_n^1(\zeta_i)}{\sum_{i=1}^N \,  \varrho_i \,  P_1^1(\zeta_i)} 
 \right)
 \, ,
\label{bratio}
\end{equation}
which separates conveniently into two independent factors: the normalized reaction factor (square braces),  
which depends on the shield/coil radius ratio $\beta$ 
(Fig.~\ref{SphericalReactionFactors});  and  the normalized free-space coefficient (round braces), which depends on the number of loops in the coil and falls off as  $N^{-2}$ for large $N$.
The combined effect of these factors is highlighted in Fig.~\ref{Bzplot} through plots of the normalized field along the $z$-axis: \begin{equation}
\frac{B_z(0,\zeta)}{B_z(0,0)} = 1 +  \sum_{n=3,5,\dots}^\infty   \!\! \frac{b_n}{b_1} \,  \zeta^{n-1}  \, .
\end{equation}
These results show that the field homogeneity of the spherical coil improves with $N$, as expected, and that it is affected by $\beta$
in a non-trivial way.

\clearpage

  \section{The solenoidal coil inside a cylindrical shield}
  \label{sec:CylindricalCurrents}

Again, we begin by considering the general azimuthal surface current $\bm F=F_\phi(z) \, \bm{\hat{\phi}}$ on a cylindrical surface of radius $a$ inside a closed cylindrical shield of radius $b$ and half-length $L$.  The resulting vector potential  is 
  \begin{equation}
A_\phi(\rho, z)= \muo \,  a \int_{-L}^LG(\rho,z;a,z') F_\phi (z') \, {\rm d}z' \, ,
\label{Aphi}
\end{equation}
where $G(\rho,z;a,z')$ is the appropriate Green function for the desired boundary condition -- magnetic~\cite{lambert} or superconducting~\cite{rigby}.  For the high-$\mu$ magnetic shield ($\mu \rightarrow \infty$)  considered here, one finds~\cite{lambert}
    \begin{eqnarray}
A_\phi&=& \frac{\muo a}{4 L} \frac{\rho^<}{\rho^>} C_0 \nonumber \\
&+&  \frac{\muo a}{L} \, \sum_{m=1}^\infty C_m \, \cos(k_e z)\, I_1(k_e \rho^<) \left[K_1(k_e \rho^>) + \frac{K_0(k_e b)}{I_0(k_e b)} I_1(k_e \rho^>) \right]  \label{Acylinder} \\  
&+& \frac{\muo a}{L} \, \sum_{m=1}^\infty D_m \, \sin(k_o z)\, I_1(k_o  \rho^<) \left[K_1(k_o  \rho^>) + \frac{K_0(k_o  b)}{I_0(k_o  b)} I_1(k_o  \rho^>) \right] \nonumber \, ,
\end{eqnarray}
where  $\rho^<$ ($\rho^>$) is the lesser (greater) of $a$ and $\rho$, 
$k_e= m \pi/L$, $k_o = (m-1/2) \pi/L$, $I_\nu$ and $K_\nu$ are the modified Bessel functions of order $\nu$, and the coefficients of the cosine and sine Fourier components are respectively
\begin{equation}
C_m = \int_{-L}^L \, {\rm d}z' \, \cos(k_e z') F_\phi(z') \, ,
\label{cn}
\end{equation}
and
\begin{equation}
D_m = \int_{-L}^L \, {\rm d}z' \, \sin(k_o  z') F_\phi(z') \, .
\label{dn}
\end{equation}
The subscripts on $k_e$ and $k_o$ indicate even ($e$) and odd ($o$) in reference to the axial symmetry of the corresponding components in Eq.~\ref{Acylinder}.
 
The resulting magnetic field is 
\begin{align}
\left(\begin{matrix}
B_\rho\\
B_z
\end{matrix}\right)
=
\frac{\mu_0C_0}{2L}
\left(\begin{matrix}
0\\
1
\end{matrix}\right)
+
\frac{\mu_0a}{L}
\sum_{m=1}^\infty{C_mk_eT(k_ea)}
\left(\begin{matrix}
\sin(k_ez)I_1(k_e\rho)\\
\cos(k_ez)I_0(k_e\rho)
\end{matrix}\right)\nonumber\\
+
\frac{\mu_0a}{L}
\sum_{m=1}^\infty{D_mk_oT(k_oa)}
\left(\begin{matrix}
-\cos(k_oz)I_1(k_o\rho)\\
\sin(k_oz)I_0(k_o\rho)
\end{matrix}\right)
\label{fullBinternal}
\end{align}
in the region $\rho<a$, and 
\begin{align}
\left(\begin{matrix}
B_\rho\\
B_z
\end{matrix}\right)
=
\frac{\mu_0a}{L}
\sum_{m=1}^\infty{C_mk_eI_1(k_ea)}
\left(\begin{matrix}
\sin(k_ez)T(k_e\rho)\\
\cos(k_ez)U(k_e\rho)
\end{matrix}\right)\nonumber\\
+
\frac{\mu_0a}{L}
\sum_{m=1}^\infty{D_mk_oI_1(k_oa)}
\left(\begin{matrix}
\cos(k_oz)T(k_o\rho)\\
\sin(k_oz)U(k_o\rho)
\end{matrix}\right)
\label{fullBexternal}
\end{align} 
in the region $a<\rho<b$, where  $T(k\xi)=K_1(k\xi) + I_1(k\xi) K_0(k b)/I_0(k b)$ and  $U(k\xi)=-K_0(k\xi) + I_0(k\xi) K_0(k b)/I_0(k b)$.
One can verify from Eqs.~\ref{fullBinternal} and~\ref{fullBexternal} that $B_\rho(\rho,\pm L)=0$ and $B_z(b,z)=0$ confirming the fact that the magnetic field must enter the high-$\mu$ shield at normal incidence.

\subsection{The  continuous solenoid}
\label{sec:ContinuousSolenoid}
A solenoid of half-length $l$ can be modelled as the continuous surface current  $F_\phi = F\, \Pi(-l,l)$, where the boxcar function $\Pi(-l,l)=1$ on the interval $-l \le z \le l$ and zero elsewhere.  From Eq.~\ref{cn}, \ref{dn}, $C_0=2Fl$, $C_m=2F  \sin(kl)/k$ for $m \ge1$, and $D_m=0$. 
Substitution into Eqs.~\ref{fullBinternal}, \ref{fullBexternal} recovers the results given by previous authors~\cite{lambert,sumner,hanson,durand}, in particular that the magnetic field in the region $\rho<a$ is 
\begin{equation}
 \left( \!\!\! 
 \begin{array}{c}
B_{\rho} \\ 
B_z
\end{array}
\!\! \! \right) 
=  \frac{\mu_o Fl}{L}
 \left(\!\! \! 
 \begin{array}{c}
0 \\ 
1 
\end{array}
\!\!\! \right) 
+  \frac{2\mu_o aF}{L} \, \sum_{m=1}^\infty  \,\sin(kl) \, T(ka)
 \left(\! \!\! 
 \begin{array}{c}
\sin(kz)\, I_1(k \rho) \\ 
\cos(kz)\, I_0(k \rho) 
\end{array}
\!\!\! \right) \, .
\label{BintContSol}
\end{equation}
In the limit  that the solenoid extends the entire  length of the shield (\textit{i.e.} $l=L$),  $C_m=0$ for all $m\ge1$ and there remains only the perfectly uniform internal field $\bm B_0=\muo F \, \bm{\hat{z}}$,  which is equivalent to that of the infinitely-long solenoid.  The interpretation, of course, is that due to boundary conditions the end faces of the high-$\mu$ shield at $z=\pm L$ act as \textit{current mirrors} making the solenoid appear infinitely long.

\subsection{Discrete Current Loops}
\label{discreteLoops}
For discrete current sources, we first consider a single loop of current $I$ located at $z=z_0$, which constitutes  a surface current $\mathbf{F_\phi}=I\delta(z-z_0)\hat{\phi}$. The coefficients of the cosine and sine Fourier components become
\begin{equation}
C_m=I\cos{(\frac{m}{L}\pi z_0)}
\label{CmLoop}
\end{equation}
and
\begin{equation}
D_m=I\sin{(\frac{m-\frac{1}{2}}{L}\pi z_0)} \, .
\label{DmLoop}
\end{equation}
When the current loop is located at the origin, $z_0=0$ and $D_m$ is zero, of course. 

Given that we are primarily concerned here with the generation of a uniform internal field, we  presently consider only a net surface current $F_\phi(z)$ that is even in $z$.  As a result,  
only the  $C_m$ terms will factor in the following analysis and for notational ease we henceforth use $k\equiv k_e=m \pi/L$.  A useful example is the  pair of current loops at $z=\pm z_0$.  The surface current here is $\mathbf{F_\phi}=I\delta(z-z_0)\hat{\phi}+I\delta(z+z_0)\hat{\phi}$, and the resulting magnetic field is
\begin{equation}
 \left( \!\!\! 
 \begin{array}{c}
B_{\rho} \\ 
B_z
\end{array}
\!\! \! \right) 
=  \frac{\mu_o I}{L}
 \left(\!\! \! 
 \begin{array}{c}
0 \\ 
1 
\end{array}
\!\!\! \right) 
+  \frac{2\mu_o aI}{L} \, \sum_{m=1}^\infty \cos(k z_0) \,k \, T(ka)
 \left(\! \!\! 
 \begin{array}{c}
\sin(kz)\, I_1(k \rho) \\ 
\cos(kz)\, I_0(k \rho) 
\end{array}
\!\!\! \right)
\label{looppairBinternal}
\end{equation}
 in the region $\rho < a$,  
and
\begin{equation}
 \left(\!\! \! 
 \begin{array}{c}
B_{\rho} \\ 
B_z
\end{array}
\!\!\! \right) 
=  \frac{2\mu_o aI}{L} \, \sum_{m=1}^\infty \cos(k z_0) \,k \, I_1(ka) \, 
 \left( \!\!\! 
 \begin{array}{c}
\sin(kz) \, T(k\rho)  \\ 
\cos(kz)\,  U(k\rho)
\end{array}
\!\!\! \right)
\label{looppairBexternal}
\end{equation}
in the region $a<\rho<b$.

\subsection{The solenoidal coil}
\label{discreteSolenoid}

\begin{figure}[htb]
\centering
\includegraphics[trim= 0 0 0 0, clip=true,keepaspectratio, width=3.2in]{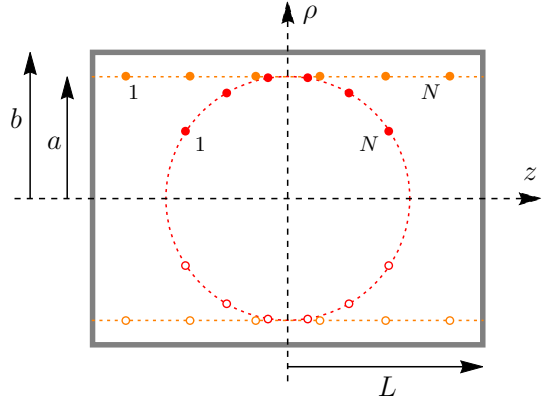}
\caption{Cross-sectional view of the $N$-loop solenoidal (yellow) and spherical  (red)  coils inside a high-$\mu$ cylindrical shield (gray). The solid versus open circles indicate current flow out of and into the page, respectively.}
\label{geometryCyl}
\end{figure}

We now seek to approximate the uniform surface current of the continuous solenoid with an evenly spaced distribution of $N$ discrete  current loops.  In particular, we consider the continuous solenoid of half-length $l=L$, which comprises a total current $I_{\rm tot} =2LF$.
To exploit the 
\textit{mirroring} effect of the end faces of the high-$\mu$ shield, the loops are separated by the distance $\delta z = 2L/N$ with the $i$-th loop  located  at $z_i=-L+\delta z(i-1/2)=L(-1 +\tfrac{2i-1}{N})$, as shown in Fig.~\ref{geometryCyl} .  Such coils have been described previously in the literature~\cite{budker}.

Each loop carries current $I=I_{\rm tot}/N$ and  
can be treated as the surface current $F_\phi(z) = I \, \delta(z-z_i)$ in cylindrical coordinates.  
Exploiting the even symmetry in $z$ (i.e., $D_m=0$), one finds  from Eqs.~\ref{fullBinternal}, \ref{fullBexternal}, and~\ref{CmLoop} that  the  field of the discretized solenoid is  
\begin{equation}
 \left( \!\!\! 
 \begin{array}{c}
B_{\rho} \\ 
B_z
\end{array}
\!\! \! \right) 
=  \frac{\mu_o NI}{2L}
 \left(\!\! \! 
 \begin{array}{c}
0 \\ 
1 
\end{array}
\!\!\! \right) 
+  \frac{\mu_o aI}{L} \, \sum_{m=1}^\infty S_N \,k \, T(ka)
 \left(\! \!\! 
 \begin{array}{c}
\sin(kz)\, I_1(k \rho) \\ 
\cos(kz)\, I_0(k \rho) 
\end{array}
\!\!\! \right) \, 
\label{BintDiscSol}
\end{equation}
in the region $\rho<a$ and 
\begin{equation}
 \left(\!\! \! 
 \begin{array}{c}
B_{\rho} \\ 
B_z
\end{array}
\!\!\! \right) 
=  \frac{\mu_o aI}{L} \, \sum_{m=1}^\infty S_N \,k \, I_1(ka) \, 
 \left( \!\!\! 
 \begin{array}{c}
\sin(kz) \, T(k\rho)  \\ 
\cos(kz)\,  U(k\rho)
\end{array}
\!\!\! \right) 
\label{BextDiscSol}
\end{equation}
in the region $a<\rho<b$.  The sum $S_N=\sum_{i=1}^N\, \cos(kz_i)$ reduces to the following  simplified form owing to the periodic structure of the coil:
\begin{eqnarray}
S_N & =&  
(-1)^m \sum_{i=1}^N\, \cos(\tfrac{m \pi}{2N}(2i-1)) \nonumber \\
& =& 
\left\{ 
\begin{array}{l l}
 N (-1)^{m+m/N} & \quad \text{if $m/N \in \mathbb{Z}^+$}     \\
 0 &   \quad \text{otherwise~\cite{TISP}.} 
\end{array}
\right.
\label{Sn_simplified}
\end{eqnarray}
In the limit $N\rightarrow \infty$, the condition that $m/N$ is a positive integer ($\mathbb{Z}^+$) implies that $S_N \rightarrow 0$ for all $m\ge 1$ and, as a result, one recovers the perfectly uniform  internal field $B_0=\muo NI/2L$ of the continuous solenoid of half-length $l=L$ with $I_{\rm tot}=NI$ and $F=NI/2L$.  Similarly, one finds via Eq.~\ref{BextDiscSol} that the field  in the region $\rho>a$ goes to zero as $N\rightarrow \infty$, as expected.

Writing Eq.~\ref{BintDiscSol} in terms of the solenoid aspect ratio $\lambda =L/a$ and the shield to solenoid radius ratio $\beta = b/a$ gives 
\begin{equation}
B_\rho(\varrho,\zeta)
=  B_0\,
\frac{2\pi}{\lambda} 
\sum_{m= N}^\infty   
m \, \tilde{S}_N \, T(\lambda,\beta)\, \sin(\tfrac{m \pi}{\lambda} \zeta)\, I_1(\tfrac{m \pi}{\lambda} \varrho)
\end{equation}

\begin{equation}
B_z(\varrho,\zeta)
= B_0 \left( 1 
+  \frac{2\pi}{\lambda} \sum_{m=N}^\infty   
m \, \tilde{S}_N   \, T(\lambda,\beta) \,
\cos(\tfrac{m\pi}{\lambda}\zeta )  \, I_0(\tfrac{m \pi}{\lambda} \varrho)
\right) \, ,
\end{equation}
where $\tilde{S}_N =S_N/N$   
and  
$T(\lambda,\beta)=K_1(\tfrac{m \pi}{\lambda})+I_1(\tfrac{m \pi}{\lambda}) K_0(\tfrac{m \pi\beta}{\lambda})/I_0(\tfrac{m \pi \beta}{\lambda})$.  The sum is restricted to integer multiples of $N$ only (\textit{i.e.,}  $m=N, 2N, 3N \dots$), since all other terms are zero by Eq.~\ref{Sn_simplified}.  The range  
 of the dimensionless variables  $\varrho=\rho/a$ and $\zeta=z/a$ is $ 0\rightarrow 1$ and $0 \rightarrow \pm \lambda$ respectively.  
 
\begin{figure}[htb]
\centering
\includegraphics[keepaspectratio, width=3.2in]{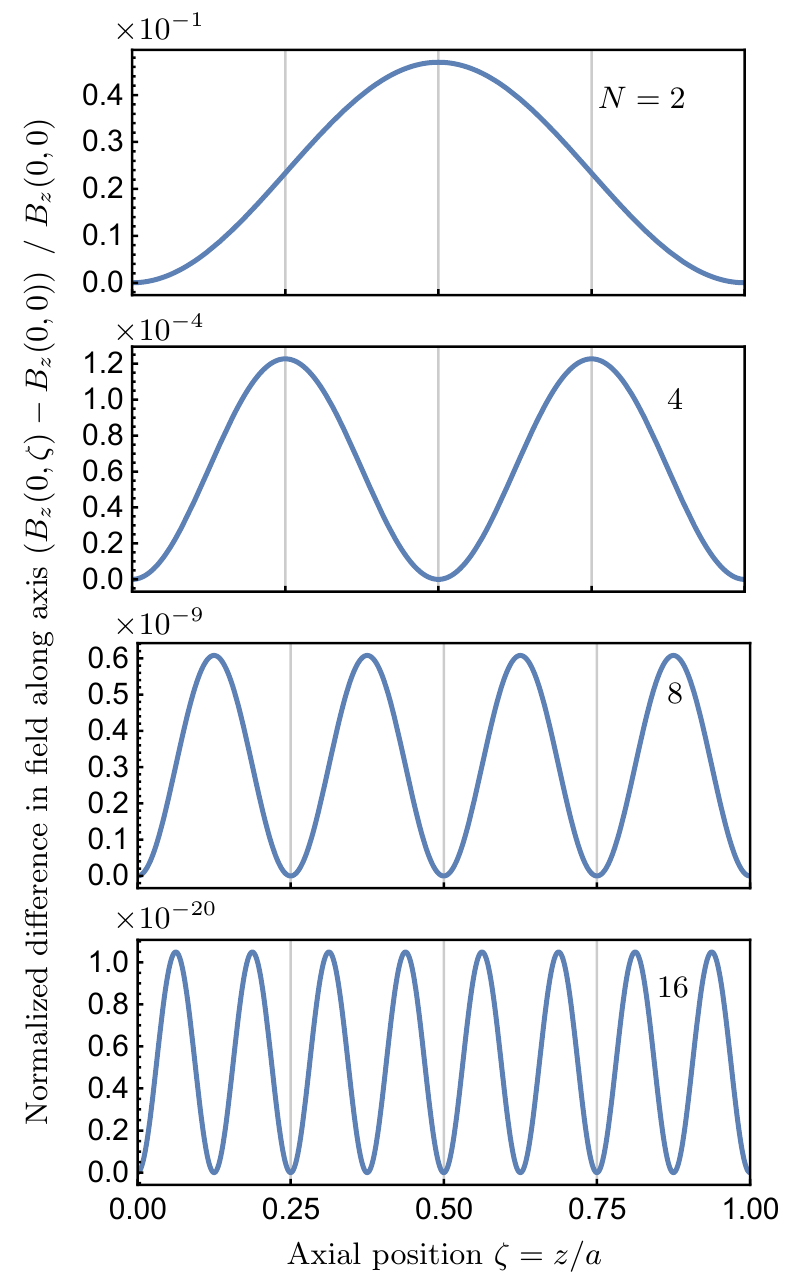}
\caption{The normalized difference of the axial field of the solenoidal coil with $\lambda=1$ and $\beta=1$.  The number of loops $N$ comprising the coil is given in the top right of each graph.  }
\label{BzSolvsN}
\end{figure}

\begin{figure}[htb]
\centering
\includegraphics[]{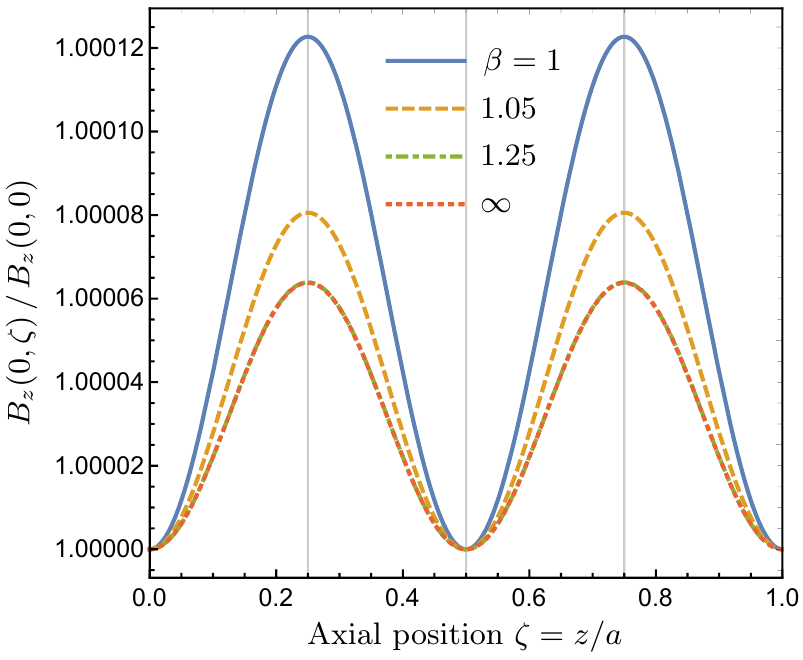}
\caption{The normalized axial field of the solenoidal coil with $N=4$, $\lambda=1$ and various $\beta$. }
\label{BzSolvsBeta}
\end{figure}

\begin{figure}[htb]
\centering
\includegraphics[]{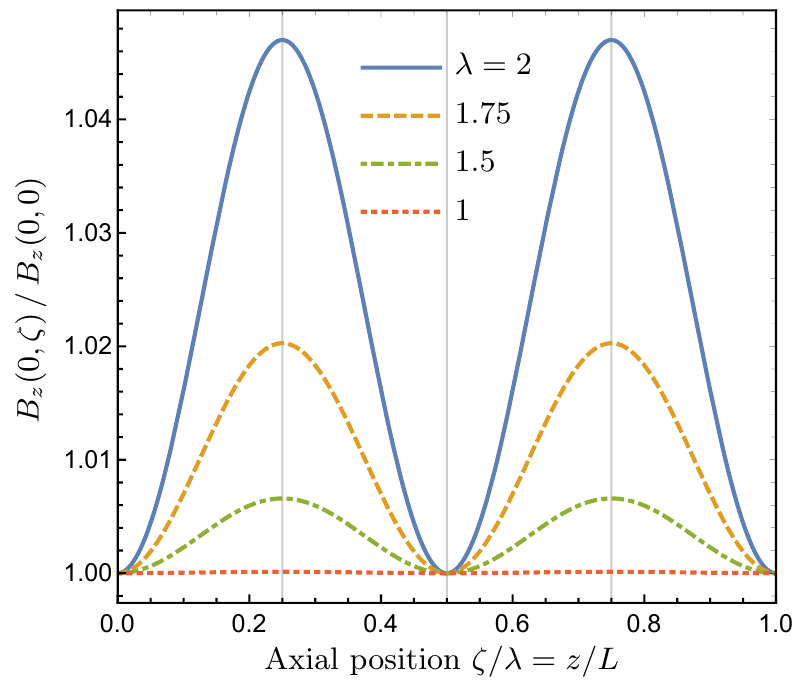}
\caption{The normalized axial field of the solenoidal coil with $N=4$, $\beta=1$ and various $\lambda$.}
\label{BzSolvsLambda}
\end{figure}

 A plot of the variation of $B_z(0,\zeta)$ normalized to the central field $B_z(0,0) = B_0 ( 1 +  \frac{2\pi}{\lambda}\, \sum_{m}^\infty   m \, \tilde{S}_N   \, T(\lambda,\beta) )$ is shown in Fig.~\ref{BzSolvsN}.  The results highlight two important differences with the spherical coil, namely that field variations are periodic over the length of the coil with a period given by $N$, and that the overall homogeneity of the field increases rapidly with $N$ owing to the behaviour of the factor $S_N$ from Eq.~\ref{Sn_simplified}.  A further difference can be seen in Fig.~\ref{BzSolvsBeta}, where the  uniformity of the discretized solenoid appears to improve uniformly as  $\beta \rightarrow \infty$. The field uniformity is also affected by the aspect ratio $\lambda$ of the coil as shown in Fig.\ref{BzSolvsLambda}.  In general, as one would expect, the longer the cylindrical shield, the more current loops are needed to achieve a desired field homogeneity.

\clearpage

Finally, we note that one can
 recaste Eq.~\ref{BintDiscSol} in terms of the polynomials $\mathcal{P}_n$ and $\mathcal{Q}_n$ via  Eqs.~\ref{BesselP1} and~\ref{BesselQ1}. 
Normalizing to the central field gives
\begin{eqnarray}
\frac{B_\rho(\varrho,\zeta)}{B_z(0,0)}& =&  \sum_{n=3,5,\dots}^\infty   \!\! \frac{b_n}{b_1}  \, \mathcal{P}_n(\varrho,\zeta) \label{BrhoSollNorm} \\
\frac{B_z(\varrho,\zeta)}{B_z(0,0)}& =& 1 +  \sum_{n=3,5,\dots}^\infty   \!\! \frac{b_n}{b_1}  \, \mathcal{Q}_n(\varrho,\zeta) \, ,
\label{BzSollNorm} 
\end{eqnarray}
where in this case 
\begin{equation}
b_1= \frac{B_z(0,0)}{B_0}=1+ \frac{2\pi}{\lambda} \sum_{m=N}^\infty  m \, \tilde{S}_N   \, T(\lambda,\beta)  \, ,
\label{b1Solenoid}
\end{equation}
\begin{equation}
b_n= 2 \left( \frac{\pi}{\lambda}\right)^n \frac{i^{(n-1)}}{(n-1)!} \sum_{m=N}^\infty  m^n \, \tilde{S}_N   \, T(\lambda,\beta)  \, .
\label{bnSolenoid}
\end{equation}
These results provide a potentially more intuitive and direct analytic comparison between the internal magnetic field of the solenoidal coil and that of   the spherical coil.  A visual comparison of Eq.~\ref{bratio} with Eqs.~\ref{b1Solenoid} and~\ref{bnSolenoid} also reveals that the reaction factor of the solenoidal is not as apparent as it is for the spherical coil.  It is possible to define such a parameter, however, as shown in~\ref{appendixB}.

\section{The spherical coil in a closed cylindrical shield}
\label{sec:SphericalCoilCylindricalShield}

For the sake of completeness, we now consider the  spherical coil inside a cylindrical shield~\cite{masuda}.  The configuration is shown in Fig.~\ref{geometryCyl}.  As discussed previously in Section~\ref{sphericalcoil}, the loops of the spherical coil are located at $z_i =a(-1 +\tfrac{2i-1}{N})$ and have corresponding radii  $\rho_i=\sqrt{a^2-z_i^2}$.  In this scenario, the net magnetic field is calculated from  Eqs.~\ref{looppairBinternal} and~\ref{looppairBexternal} of Section~\ref{discreteLoops} with $a$  replaced by $\rho_i$ for each pair of loops in the coil, and  keeping in mind that the choice of formula depends on whether the field point $\rho$ is less than or greater than  $\rho_i$.  Also, if $N$ is odd, one must divide the appropriate formula by a factor~2 for the single loop at $z_i=0$. 

Owing to the broken symmetry here -- namely, a spherical current distribution inside a cylindrical magnetic structure -- this configuration does not  produce a perfectly uniform magnetic field  as $N \rightarrow \infty$.  In this limit, the internal field ($r<a$) is the superposition of the uniform internal field of the sine-theta surface current distribution plus the response of the cylindrical shield to the external dipole field of the sine-theta surface current distribution.     As might be expected, then, the spherical coil in the cylindrical shield is least homogeneous of the three coil types explored in this paper, as shown in the next section.

\section{Comparison of all coil types}
\label{sec:Comparison}
With the analysis of Sections~\ref{Sec:SphericalCurrents} to~\ref{sec:SphericalCoilCylindricalShield} complete, one can now use these results to  compare coil designs and search for optimal solutions.   Parameter space is large, however, and design constraints are highly dependent on the specific application.  As a result, an exhaustive study of any kind is beyond the scope of this work.  Rather, we provide a simple comparison that highlights above all else that the solenoidal coil in the cylindrical shield can achieve excellent field homogeneity over much larger volumes than the other two coil types.

\begin{figure}[htb]
\centering
\includegraphics[]{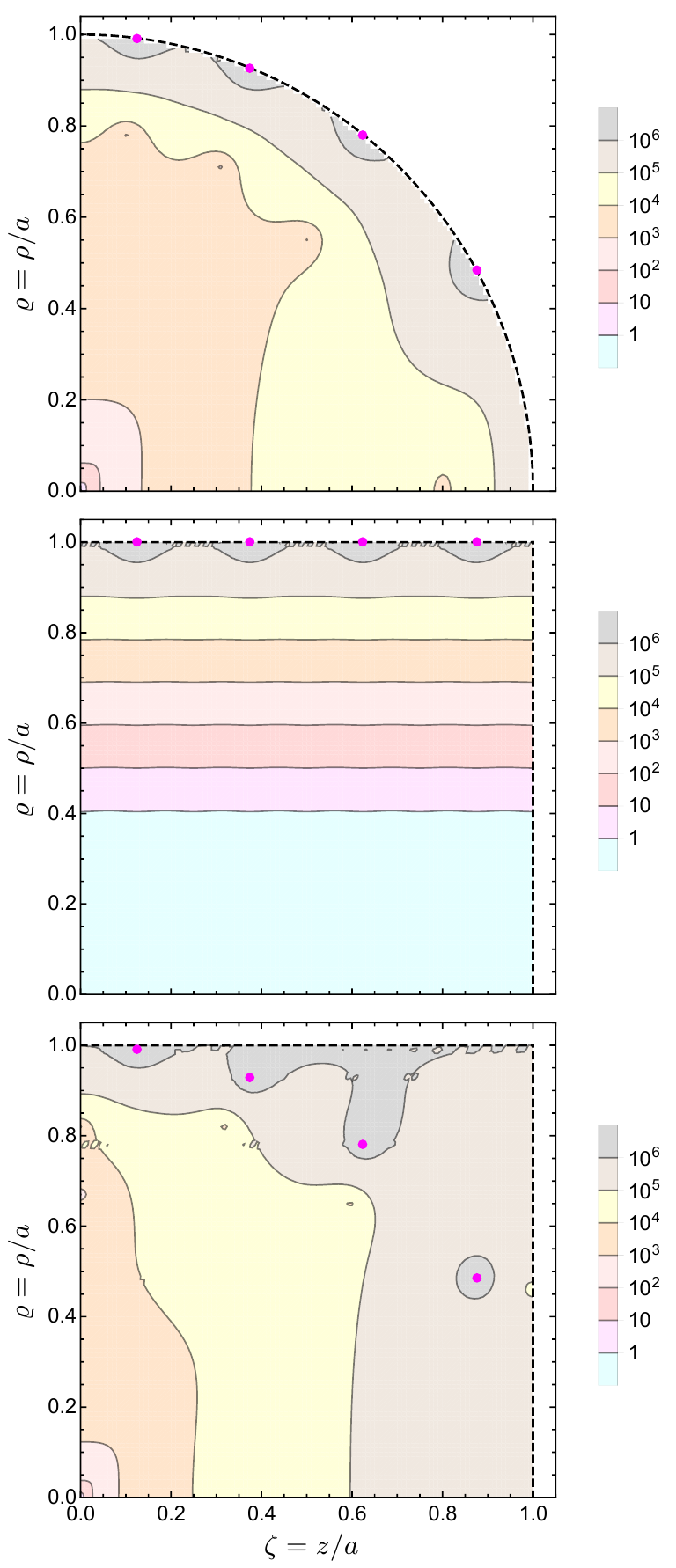}
\caption{Contour maps of $\Delta B$ for the spherical coil in a spherical shield (top), the solenoidal coil in a cylindrical shield (middle), and the spherical coil in a cylindrical shell (bottom). The contour values are in parts per million. The magenta dots are the current loops, and the dashed black lines indicate the inner boundary of the shield. 
}
\label{uniformitymaps}
\end{figure}

For this comparison we purposely chose a rather modest number of loops ($N=8$), in order to demonstrate that the homogeneity of the solenoidal coil in the cylindrical shield improves rapidly with $N$.  We also chose tight-fitting coils (i.e., $b=a$ and $L=a$), which is not optimal  for any of the three coil types in terms of field homogeneity but can be advantageous in all cases in that it maximizes experimental space.   The results are shown in Fig.~\ref{uniformitymaps}  ~as contour maps of normalized deviation from the central field defined as 
\begin{equation}
\Delta B(\rho,z)=\frac{\sqrt{B_\rho(\rho,z)^2+(B_z(\rho,z)-B_z(0,0))^2}}{B_z(0,0)} \, .
\label{DeltaB}
\end{equation}
Because of the axial symmetry here, $B_\rho(0,z)=0$ along the $z$-axis, and this definition is consistent with the way results are presented in Sections~\ref{Sec:SphericalCurrents} and~\ref{sec:CylindricalCurrents}.  Here, one sees that the spherical coil, whether in a spherical shield or a cylindrical shield, has relatively poor homogeneity for so few loops. In contrast, the solenoidal coil in the cylindrical shield has a notably large region of homogeneity, with $\Delta B(\rho,z)$ within 1 part per million over $16\%$ of the coil volume.

\clearpage

\section{Analysis of the impact of winding misplacement} 
\label{sec:WindingMisplacement}

In previous sections, it was assumed that the current loops, or coil windings, were perfectly located relative to one another and the shield as given by the formulae for $(\rho_i, z_i)$.  This section provides a sample analysis of how departures from these locations impacts the homogeneity of the coil.  We focus on the solenoidal coil here since it has the potential to produce the most uniform field among the three coil types.  We further choose to explore two specific winding misplacements: a global translation $d$ of all loops along the $z$-direction, and 
a global scaling of the coil in the $z$-direction such that its half-length is no longer that of the shield, \textit{i.e.} $l \neq L$.  To include these effects, we generalize the axial location of the loops given in Sec.~\ref{discreteSolenoid} to $z_i=d+l(\frac{2i-1}{N}-1)$, which results in the following Fourier components: 
\begin{equation}
C_m=I\sum_{i=1}^N{\cos\left[k_e\left(d+l\left(\frac{2i-1}{N}-1\right)\right)\right]}
\end{equation}
\begin{equation}
D_m=I\sum_{i=1}^N{\sin\left[k_o\left(d+l\left(\frac{2i-1}{N}-1\right)\right)\right]} \, .
\end{equation}
One must use Eqns.~\ref{fullBinternal} and ~\ref{fullBexternal} now to calculate the magnetic field generated by the coil. 

\begin{figure}[htb]
\centering
\includegraphics[]{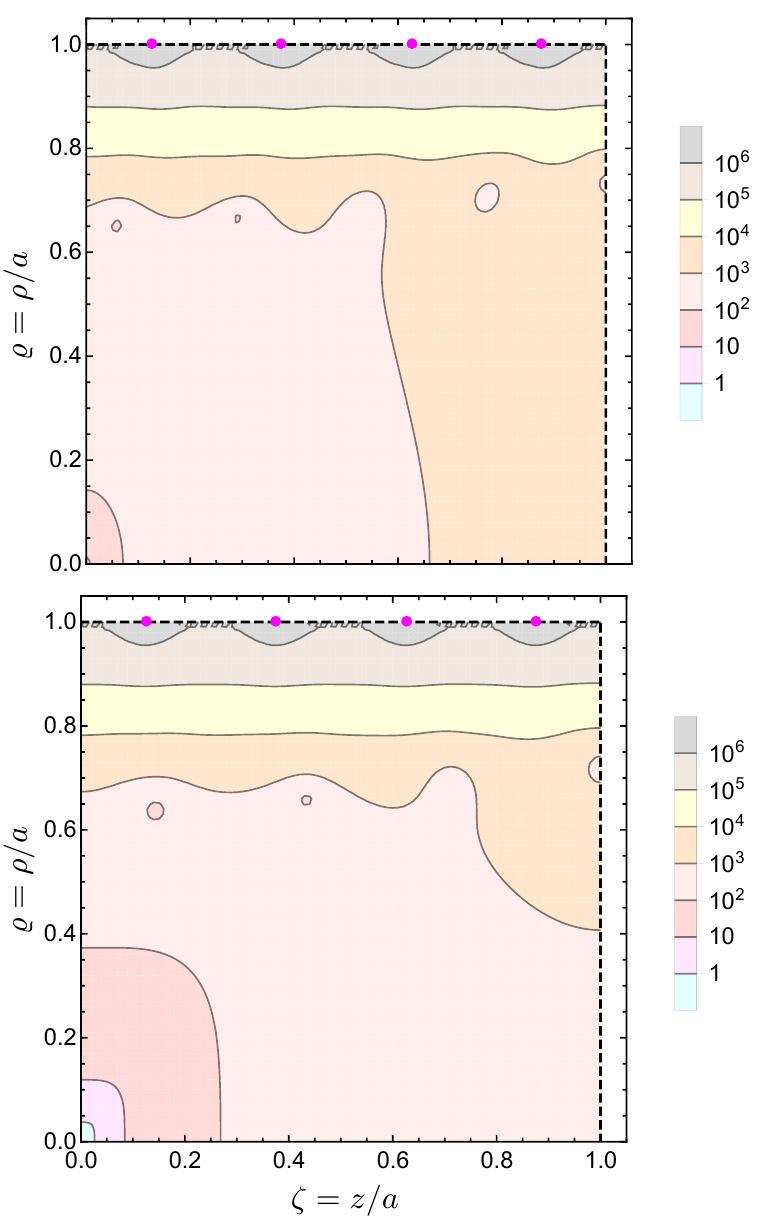}
\caption{Contour maps of $\Delta B$ for the solenoidal coil of Fig.~\ref{deformedcoilmap} with $d/L=0.001$ and $l/L=1.000$ (top), and $d/L=0.000$ and $l/L=1.001$ (bottom).  As above, the contours are in parts per million, the magenta dots are the current loops, and the dashed black lines indicate the inner boundary of the shield.}
\label{deformedcoilmap}
\end{figure}

\begin{figure}[htb]
\centering
\includegraphics[]{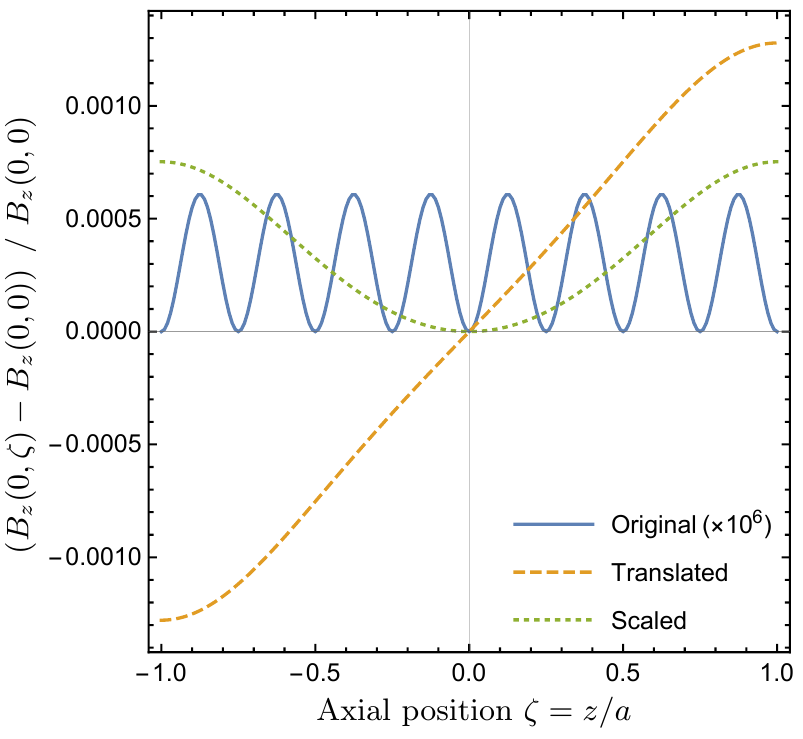}
\caption{The normalized difference of the axial field of the solenoidal coil of Fig.~\ref{deformedcoilmap}
that has been either translated ($d=0.001L$, $l=L$) or scaled ($d=0$, $l=1.001L$).  
The result for the original coil (without any translation or scaling) is multiplied by $10^6$ for clarity. 
}
\label{deformedcoilbzprofile}
\end{figure}

As an example, we consider the same solenoidal coil of Fig.~\ref{uniformitymaps} but with $d/L=0.001$ and $l/L=0.001$.   This equates to a 1-mm winding misplacement, due to translation or scaling, for a coil of  $L=1$~m, and as such represents a reasonable scenario that one might encounter for a typical sized EDM experiment. Figure~\ref{deformedcoilmap} shows contour maps of  $\Delta B(\rho,z)$, while Fig.~\ref{deformedcoilbzprofile} shows the normalized field difference along the central axis.  The homogeneity of the field is dramatically reduced compared to that in Fig.~\ref{uniformitymaps}.  Moreover, one finds that the introduction of a translation results in dominant first-order gradient near the center of the coil, while a scaling results in a dominant second-order gradient.  An analysis such as this can be useful for setting construction tolerances on solenoidal (or similar) coils, as well as for the design of shim coils that can be used to correct field inhomogeneities.

\clearpage

\section{Conclusion}

Precision EDM experiments  require electromagnetic coils that can generate highly homogeneous magnetic fields within a shielded volume.   The design of such coils is assisted by analytic models that can serve as theoretical and conceptual guides, as well as  benchmarks for finite-element analysis.  In this paper, we presented complete solutions for three such models: the spherical coil in a spherical shield, the solenoidal coil in a cylindrical shield, and the spherical coil in the cylindrical shield.  The method of solution is readily generalized to any axisymmetric current distribution within a highly-permeable spherical or cylindrical shield.  This  was used here  to explore field inhomogeneities related to winding misplacement.  A key finding of this work is that the solenoidal coil in a cylindrical shield can achieve very homogeneous fields over  large volumes with even a fairly small number of current loops.  This, along with its more practical geometry, makes it the most attractive design of the three explored here.

\section{Acknowledgements}

The authors gratefully acknowledge the support of the Natural Sciences and Engineering Research Council of Canada as well as Osaka University for a visiting professorship for CPB.

\appendix
\section{The Polynomials $\mathcal{P}_n$ and $\mathcal{Q}_n$ }
\label{appendixA}
 
Transforming the   internal field of the spherical coil (Eq.~\ref{BNloop}) to cylindrical components results in the following spatially dependent factors that appear in $B_\rho$ (Eq.~\ref{BrhoSpherical}) and $B_z$ (Eq.~\ref{BzSpherical}):
\begin{eqnarray}
&&r^{n-1}\left( \sin\theta \, P_n(\cos\theta)-\cos\theta \, \tfrac{1}{n}\,P_n^1(\cos\theta) \right) \nonumber \\ 
&&~~~~=r^{n-1}  \left(P_n(\cos\theta)-\cos\theta \, P_{n-1}(\cos\theta) \right)/ \sin\theta  \nonumber \\
&&~~~~=- \tfrac{1}{n} \,r^{n-1}  \,\sin\theta \, P_{n-1}'(\cos\theta)\nonumber \, \\
&&~~~~=- \tfrac{1}{n} \,r^{n-1} \, P_{n-1}^1(\cos\theta) \, ,
 \label{polyP} 
\\[4pt]
 &&r^{n-1} \left( \cos\theta \, P_n(\cos\theta)+\sin\theta \, \tfrac{1}{n}\,P_n^1(\cos\theta) \right) \nonumber \\ 
  &&~~~~= r^{n-1}  \, P_{n-1}(\cos\theta)  \, ,
 \label{polyQ}
\end{eqnarray}
 Equations~\ref{polyP} and~\ref{polyQ} are used to define  
   $\mathcal{P}_n(\rho,z)$ and $\mathcal{Q}_n(\rho,z)$ respectively,  
 with $\cos\theta =z/r$, $\sin\theta=\rho/r$, and $r^2=\rho^2+z^2$.  
 The  first few polynomials  are given in Table~\ref{poly}.  Dividing these by $a^{n-1}$ gives the dimensionless polynomials $\mathcal{P}_n(\varrho,\zeta)$and $\mathcal{Q}_n(\varrho,\zeta)$ appearing in the main text. 

\begin{table}[h]
\begin{center}
\begin{tabular}{c  c c  }
\toprule
$n$ & $\mathcal{P}_n(\rho,z)$ & $\mathcal{Q}_n(\rho,z)$ \\
\midrule
\noalign{\smallskip}
1 	& 0		& 1  \\
\noalign{\smallskip}
\midrule
\noalign{\smallskip}
2	& $-\rho$ 		&  $z$  \\
\noalign{\smallskip}
\midrule
\noalign{\smallskip}
3	& $-\rho z$ 		&  $-\tfrac{1}{2}( \rho^2 - 2   z^2)$  \\
\noalign{\smallskip}
\midrule
\noalign{\smallskip}
4	& $-\tfrac{3}{8}(4\rho z^2 - \rho^3)$ 		&  $-\tfrac{1}{2}( 3z\rho^2 - 2   z^3)$  \\
\noalign{\smallskip}
\midrule
\noalign{\smallskip}
5	&$\tfrac{1}{2}(3 \rho^3 z - 4 \rho  z^3)$ 		 & $\tfrac{1}{8}(3 \rho^4  - 24 \rho^2  z^2+ 8 z^4)$  \\
\noalign{\smallskip}
\midrule
\noalign{\smallskip}
6	& $-\tfrac{5}{16}(\rho^5 - 12\rho^3z^2 + 8\rho z^4)$ 		&  $\tfrac{1}{8}( 15\rho^4z - 40\rho^2z^3 + 8z^5)$  \\
\noalign{\smallskip}
\midrule
\noalign{\smallskip}
7	&$-\tfrac{3}{8} \left(5 \rho ^5 z-20 \rho ^3 z^3+8 \rho  z^5\right)$ & $-\tfrac{1}{16}(5 \rho^6  - 90 \rho^4  z^2 +120\rho^2z^4- 16z^6)$ \\
\noalign{\smallskip}
\bottomrule
\end{tabular}
\end{center}
\caption{The first seven degrees of the polynomials $\mathcal{P}_n(\rho,z)$ and $\mathcal{Q}_n(\rho,z)$.}
\label{poly}
\end{table}

Given that the spherical and  solenoidal coils are both axisymmetric, it is not surprising that there exists a fairly simple relationship between the polynomials $\mathcal{P}_n(\rho,z)$ and $\mathcal{Q}_n(\rho,z)$  and the products of trigonometric and modified Bessel functions that appear in Section~\ref{sec:CylindricalCurrents}.  
For example, using the series expansions of $\sin(kz)$ and $I_1(k\rho)$~\cite{TISP} 
one finds that
\begin{eqnarray}
\sin(kz) \, I_1(k\rho)&=&\frac{k^2}{2} \rho z +\frac{k^4}{48}(3 \rho^3 z - 4 \rho  z^3) \nonumber \\
&&+\frac{k^6}{1920}(5 \rho ^5 z-20 \rho ^3 z^3+8 \rho  z^5) +\dots \nonumber \\[4pt]
&=& -\frac{k^2}{2} \mathcal{P}_3(\rho,z)+\frac{k^4}{24} \mathcal{P}_5(\rho,z)-\frac{k^6}{720} \mathcal{P}_7(\rho,z)+\dots \nonumber \\[4pt]
&=&\sum_{n=3,5,\dots}^\infty \!\! \frac{(ik)^{n-1}}{(n-1)!} \, \mathcal{P}_n(\rho,z) \, ,
 \label{BesselP1}
\end{eqnarray}
where $i=\sqrt{-1}$. Similarly, 
\begin{equation}
\cos(kz) \, I_0(k\rho) = \sum_{n=1,3,5,\dots}^\infty \!\! \frac{(ik)^{n-1}}{(n-1)!} \, \mathcal{Q}_n(\rho,z) \, ,
 \label{BesselQ1}
 \end{equation}
\begin{equation}
\cos(kz) \, I_1(k\rho) = \sum_{n=2,4,6,\dots}^\infty \!\! \frac{i^nk^{n-1}}{(n-1)!} \, \mathcal{P}_n(\rho,z) \, ,
 \label{BesselP2}
\end{equation}
\begin{equation}
\sin(kz) \, I_0(k\rho) = \sum_{n=2,4,6,\dots}^\infty \!\! \frac{-i^nk^{n-1}}{(n-1)!} \, \mathcal{Q}_n(\rho,z) \, .
 \label{BesselQ2}
\end{equation}

 \section{The reaction factor of the solenoidal coil}
\label{appendixB}
 Using the methods of Refs.~\cite{turner, cb1}, the vector potential of a  continuous solenoid  in  free-space  is
 \begin{equation}
A_\phi
=  \frac{2\mu_o a F_\phi}{\pi} \, \int_0^\infty {\rm d}k \,\cos kz \, \frac{\sin kl}{k} \, I_1(k\rho^<)K_1(k\rho^>) \, ,
\label{ASolFree}
\end{equation}
where $k$ is a continuous variable, and $\rho^<$ ($\rho^>$) is the lesser (greater) of $a$ and $\rho$.
For the discrete current loop at $z=z_i$ in free space, the corresponding vector potential is 
 \begin{equation}
A_\phi
=  \frac{\mu_o I a}{\pi} \, \int_0^\infty {\rm d}k \,\cos k(z-z_i) \, I_1(k\rho^<)K_1(k\rho^>) \, ,
\label{AloopFree}
\end{equation}
 which agrees with Jackson~\cite{jackson}.  The resulting magnetic field is 
 \begin{equation}
 \left(\!\! \! 
 \begin{array}{c}
B_{\rho} \\ 
B_z
\end{array}
\!\!\! \right) 
=  \frac{\mu_o I a}{\pi} \, \int_0^\infty {\rm d}k \,k \, K_1(ka) \, 
 \left( \!\!\! 
 \begin{array}{c}
 \sin k(z-z_i) \, I_1(k \rho) \\ 
 \cos k(z-z_i) \,  I_0(k\rho)
\end{array}
\!\!\! \right) 
\end{equation}
in the region  $\rho<a$, and
 \begin{equation}
 \left(\!\! \! 
 \begin{array}{c}
B_{\rho} \\ 
B_z
\end{array}
\!\!\! \right) 
=  \frac{\mu_o I a}{\pi} \, \int_0^\infty {\rm d}k \,k \, I_1(ka) \, 
 \left( \!\!\! 
 \begin{array}{c}
 \sin k(z-z_i) \, K_1(k \rho) \\ 
 -\cos k(z-z_i) \,  K_0(k\rho)
\end{array}
\!\!\! \right) \, 
\end{equation}
in the region  $\rho>a$
 
 When summing over the $N$ loops of the discretized solenoidal coil, the internal field becomes
 \begin{equation}
 \left(\!\! \! 
 \begin{array}{c}
B_{\rho} \\ 
B_z
\end{array}
\!\!\! \right) 
=  \frac{\mu_o I a}{\pi} \!\int_0^\infty {\rm d}k \,k \, K_1(ka) \, \left( \sum_{i=1}^N\, \cos(kz_i) \right) 
 \left( \!\!\! 
 \begin{array}{c}
 \sin kz \,\, I_1(k \rho) \\ 
 \cos kz \,\,  I_0(k\rho)
\end{array}
\!\!\! \right) \, .
\end{equation}
 where  $z_i=L(-1 +\tfrac{2i-1}{N})$.
 These equations can be expanded using Eqs.~\ref{BesselP1} and~\ref{BesselQ2} to give
 \begin{eqnarray}
B_\rho(\varrho,\zeta)& =& \frac{\mu_o I }{\pi a} \sum_{n=3,5,\dots}^\infty   \!\! b_n^{\rm free}   \, \mathcal{P}_n(\varrho,\zeta) \label{BrhoSolfree} \\
B_z(\varrho,\zeta)& =& \frac{\mu_o I}{\pi a}\sum_{n=1,3,5,\dots}^\infty   \!\! b_n^{\rm free}  \, \mathcal{Q}_n(\varrho,\zeta) \, ,
\label{BzSolfree} 
\end{eqnarray}
 where
 \begin{equation}
b_{n}^{\rm free} = \frac{(i)^{n-1}}{(n-1)!} \int_0^\infty {\rm d}\kappa \,\kappa^n  \,  K_1(\kappa) \, \left( \sum_{i=1}^N\, \cos(\kappa \zeta_i) \right)\,
\label{bnfree}
\end{equation}
is the free-space coefficient, $\kappa =ka$, and $\zeta_i=\lambda(-1 +\tfrac{2i-1}{N})$.
From a comparison of Eqs.~\ref{BrhoSollNorm}--\ref{BzSollNorm} with Eqs.~\ref{BrhoSolfree}--\ref{BzSolfree}, the reaction factor for the $n^{\rm th}$ term of the field expansion of the  solenoidal coil in a cyindrical shield is $(N \pi b_n)/ (2 \lambda b_{n}^{\rm free})$, which depends on  $N$, $\beta$ and $\lambda$.
 As $N\rightarrow \infty$, the reaction factor for the uniform field term ($n=1$) tends to a constant greater than one, while for all higher order terms ($n=3,5,7 \dots$) the reaction factor goes to zero.

\end{document}